

\input harvmac
\def\rhob{{\rho\kern-0.465em \rho}}

\def\ontopss#1#2#3#4{\raise#4ex \hbox{#1}\mkern-#3mu {#2}}

\setbox\strutbox=\hbox{\vrule height12pt depth5pt width0pt}
\def\tablerule{\noalign{\hrule}}

\def\strut{\relax\ifmmode\copy\strutbox\else\unhcopy\strutbox\fi}

\nref\rvgr{G. von Gehlen and V. Rittenberg, $Z_n$-symmetric quantum
chain with an infinite set of conserved charges and $Z_n$ zero modes,
 Nucl. Phys. B257[FS14]
(1985) 351.}
\nref\rhkd{S. Howes, L.P. Kadanoff and M. den Nijs, Quantum model for
commensurate-incommensurate transitions, Nucl. Phys
B215[FS7] (1983) 169.}
\nref\rons{L. Onsager, Crystal statistics I. A two-dimensional model
with an order disorder transition, Phys. Rev. 65 (1944) 117.}
\nref\rampty{H. Au-Yang, B.M. McCoy, J.H.H. Perk, S. Tang and M-L
Yan, Commuting transfer matrices in the chiral Potts models: Solutions
of the star-triangle equations,  Phys. Lett. A123 (1987) 138.}
\nref\rbpa{R.J. Baxter, J.H.H. Perk and H. Au-Yang, New solutions of
the star-triangle equations for th chiral Potts model, Phys. Lett. A128
(1988) 138.}
\nref\rayp{H. Au-Yang and J.H.H. Perk, Onsager's star-triangle
equation: Master key to integrability, Adv. Stud. in Pure Math. 19
(1989) 57.}
\nref\rampt{G. Albertini, B.M. McCoy, J.H.H. Perk and S. Tang,
Excitation spectrum and order parameter for the integrable $N$-state
chiral Potts model, Nucl. Phys. B314 (1989) 741.}
\nref\rbax{R.J. Baxter, The superintegrable chiral Potts model,
 Phys. Letts. A133 (1988) 185.}
\nref\ramp{G. Albertini, B.M. McCoy and J.H.H. Perk, Eigenvalue
spectrum of the superintegrable chiral Potts model, Adv. Stud. in
Pure Math. 19 (1989) 1.}
\nref\rdav{B. Davies, Onsager's algebra and superintegrability, J.
Phys. A23 (1990) 2245; Onsager's algebra and the Dolan-Grady
condition in the non-self-dual case, J. Math. Phys. 32
(1991) 2945.}
\nref\rdkma{S. Dasmahapatra, R. Kedem and B.M. McCoy, Spectrum and
completeness of the three-state superintegrable chiral Potts model,
 Nucl. Phys. B396
(1993) 506.}
\nref\rfz{V.A. Fatteev and A.B. Zamolodchikov, Self-dual solutions of
the star-triangle relations in $Z_N$ models, Phys. Letts. A92 (1982)
37.}
\nref\rbs{V.V. Bazhanov and Ya.G. Stroganov, Chiral Potts model as a
descendent of the six vertex model, J. Stat. Phys. 59 (1990) 799.}
\nref\rbbp{R.J. Baxter, V.V. Bazhanov and J.H.H. Perk, Functional
relations for transfer matrices of the chiral Potts model, Int. J.
Mod. Phys. 5 (1990) 803.}
\nref\rbaxb{R.J. Baxter,  Chiral Potts model: eigenvalues of the
transfer matrix, Phys. Letts. A146 (1990) 110.}
\nref\radma{G. Albertini, S. Dasmahapatra, and B.M. McCoy, Spectrum
and completeness of the integrable 3-state Potts model: a finite size
study,  Int. J. Mod. Phys. A7 (Suppl. 1A) (1992) 1.}
\nref\radmb{G. Albertini, S. Dasmahapatra and B.M. McCoy, Spectrum
doubling and the extended Brillouin zone in the excitations of the
three-state Potts spin chain,  Phys.
Letts. A 170 (1992) 397.}
\nref\rmr{B.M. McCoy and S-S. Roan, Excitation spectrum and phase
structure of the chiral Potts model,  Phys. Letts. A 150 (1990) 347.}
\nref\rvoge{G. von Gehlen, Phase diagram and two-particle structure
of the $Z_3$-chiral Potts model, hep-th 9207022, to be published in
proceedings of {\it Int. Symposium on Advanced Topics of Quantum
Physics,} Taiyuan, China.}
\nref\rvgho{G. von Gehlen and A. Honecker, Multiparticle structure in
the $Z_n$-chiral Potts models, J. Phys. A26 (1993) 1275.}
\nref\rhaho{N.S. Han and Z. Honecker, Low-temperature expansions and
correlation functions of the $Z_3$-chiral Potts model, hep-th/9304083}
\nref\rdkmm{S. Dasmahapatra, R. Kedem, B.M. McCoy and E. Melzer,
Virasoro characters from the Bethe equations for the critical
ferromagnetic three-state Potts model, J. Stat. Phys. 74 (1994) 239.}
\nref\rcardy{J. Cardy, Effect of boundary conditions on the operator
content of two-dimensional conformal field theories,
 Nucl. Phys. B 275[FS17] (1986) 200.}
\nref\rfg{P. Fendley and P. Ginsparg, Non-critical orbifolds,
Nucl. Phys. B 324 (1989) 549.}
\nref\rro{Ph. Roche, Ocneanu cell calculus and integrable lattice
models, Comm. Math. Phys. 127 (1990) 395.}
\nref\rpz{P.A. Pearce and Y-K. Zhou, Intertwiners and the A-D-E
lattice models, Int. Jour. of Mod. Phys. B7 (1993) 3649.}
\nref\rbaza{V.V. Bazhanov and N. Yu. Reshetikhin, Restricted solid on
solid models connected with simply laced algebras and conformal field
theory, J. Phys. A 23 (1990) 1477.}
\nref\rbazb{V.V. Bazhanov and N.Yu. Reshetikhin, Scattering amplitudes
in offcritical models and RSOS integrable models, Prog. Theo. Phys.
Suppl. 102 (1990) 301.}
\nref\rkun{A. Kuniba, Thermodynamics of the $U_q(X_r^{1})$ Bethe ansatz system
with $q$ a root of unity, Nucl. Phys. B389 (1993) 209.}
\nref\rhexa{R.J. Baxter and P.A. Pearce, Hard hexagons: Interfacial
tension and correlation length, J. Phys. A 15 (1982) 897.}
\nref\rhexb{V.V. Bazhanov and N.Yu. Reshetikhin, Critical RSOS models
and conformal field theory, Int. J. Mod. Phys. A 4 (1989) 115.}
\nref\rhexc{A. Kl{\" u}mper and P.A. Pearce, Analytic calculation of
scaling dimensions of tricritical hard squares and critical hard
hexagons, J. Stat. Phys. 64 (1991) 13.}
\nref\rlepr{J. Lepowsky and M. Primc, Structure of the Standard
Modules for the Affine Lie Algebra $A^{(1)}_1,$ Cont. Math., Vol. 46
(AMS, Providence, Rhode Island, 1985).}
\nref\rberk{A. Berkovich, Fermionic counting of RSOS-states and
Virasoro character formulas for the unitary minimal series $M(\nu,\nu
+1).$ Exact results, preprint hep-th 9403073.}
\nref\rfadtak{L.D. Faddeev and L.A. Takhtajan, What is the spin of a
spin wave?, Phys. Lett. 85A (1981) 375.}
\nref\rfadtakb{L.D. Faddeev and L.A. Takhtajan, Spectum and scattering
of excitations in the one-dimensional isotropic Heisenberg model,
J. of Sov. Math. 24 (1984) 241.}
\nref\rtaksuz{M. Takahashi and M. Suzuki, One-dimensional anisotropic
Heisenberg model at finite temperatures, Prog. Theo. Phys. 48 (1972)
2187.}
\nref\rgau{M. Gaudin, Thermodynamics if the Heisenberg-Ising ring for
$\delta \geq 1.$ Phys. Rev. Letts. 26 (1971) 1301.}
\nref\rkor{V.E. Korepin, Direct calculation of the S matrix of the
massive Thirring model, Theoretical and Mathematical Physics 41 (1979) 953. }

\nref\rhal{F.D.M. Haldane, ``spinon-gas'' description of the $S=1/2$
Heisenberg chain with inverse square exchange: exact spectrum and
thermodynamics, Phys. Rev. Letts. 66 (1991) 1529.}
\nref\rhalb{F.D.M. Haldane, Z.N.C. Ha, J.C. Talstra, D. Bernard and
V. Pasquier, Yangian symmetry of integrable quantum chains with
long-range interactions and a new description of states in conformal
field theory, Phys. Rev. Letts. 69 (1992) 2021.}
\nref\rha{Z.N.C. Ha and F.D.M. Haldane, Squeezed strings and Yangian
symmetry of the Heisenberg chain with long-range interactions,
Phys. Rev. B 47 (1993) 12459.}
\nref\rbps{D. Bernard, V. Pasquier and D. Serban, Spinons in conformal
field theory, preprint SPhT/94/039.}
\nref\rsch{P. Bouwknegt, A.W.W. Ludwig and K. Schoutens, Spinon
bases, Yangian Symmetry and Fermionic representations of characters in
conformal field theory, preprint PUPT-1469, hep-th/9406020}
\nref\rchara{R. Kedem, T.R. Klassen, B.M. McCoy and E. Melzer,
Fermionic sum representations for conformal field theory characters,
Phys. Letts.B 307 (1993) 68.}
\nref\rcharb{S. Dasmahapatra, R. Kedem, T.R. Klassen, B.M. McCoy and
E. Melzer, Quasi-particles, conformal field theory and $q$ series, Int.
J. Mod. Phys. B7 (1993) 3617.}
\nref\rkm{M. Kashiwara and T. Miwa, A class of elliptic solutions to
the star triangle equations, Nucl. Phys. B275[FS17] (1986) 121.}
\nref\rpas{V. Pasquier, $D_n$ models: local densities, J. Phys. A 20
(1987) L221.}

\Title{\vbox{\baselineskip12pt\hbox{ITPSB 94-013}
\hbox{HEP-TH 9405089}}}
{\vbox{\centerline{Quasi-particles in the Chiral Potts model}}}
\centerline{Rinat Kedem~\foot{rinat@kurims.kyoto-u.ac.jp}}

\bigskip\centerline{\it Research Institute for Mathematical Sciences}
\centerline{\it Kyoto University}
\centerline{\it Kyoto 606-01, Japan}
\bigskip
\centerline{and}
\bigskip
\centerline{ Barry~M.~McCoy~\foot{mccoy@max.physics.sunysb.edu}}

\bigskip\centerline{\it Institute for Theoretical Physics}
\centerline{\it State University of New York}
\centerline{\it  Stony Brook,  NY 11794-3840}

\vskip 10mm

\centerline{\bf Abstract}

We study the quasi-particle spectrum of the
integrable three-state chiral Potts chain in the massive phase
by combining a numerical
study of the zeroes of associated transfer
matrix eigenvalues with the exact results of the
ferromagnetic three-state Potts
chain and the three-state superintegrable chiral Potts model. We
find that the spectrum is described in terms of
quasi-particles with momenta restricted only to
segments of the Brillouin zone $0\leq P \leq 2\pi$ where the
boundaries of the segments depend on the chiral angles of the model.

\vskip 3mm

\Date{\hfill 5/94}
\vfill\eject

\newsec{Introduction}

In 1985, von Gehlen and Rittenberg~\rvgr~made a profound discovery
concerning the $Z_n$ symmetric chiral Potts spin chain~\rhkd~
with the Hamiltonian
\eqn\cp{H_{{\rm CP}}=A_0+kA_1~,}
where
\eqn\azero{A_0=-\sum_{j=1}^L\sum_{n=1}^{N-1}{e^{i(2n-N)\phi/N}\over
\sin(\pi n/N)}(Z_jZ^{\dagger}_{j+1})^n~,\qquad
A_1=-\sum_{j=1}^L\sum_{n=1}^{N-1}{e^{i(2n-N){\bar
\phi}/N}\over \sin(\pi n/N)}X_j^n~.}
The matrices $X_j,Z_j$ are defined by
\eqn\exj{X_j=I_N\otimes\cdots\underbrace{X}_{{\rm
site}~j}\cdots\otimes I_N~,\qquad
Z_j=I_N\otimes\cdots\underbrace{Z}_{{\rm
site}~j}\cdots\otimes I_N~,}
where $I_N$ is the $N\times N$ identity matrix, the elements of the $N\times
N$ matrices $X$ and $Z$ are
\eqn\exel{X_{l,m}=\delta_{l,m+1}~({\rm mod}~N)~,\qquad
Z_{l,m}=\delta_{l,m}\omega^{l-1}}
and $\omega=e^{2\pi i/N}.$
They demonstrated that in the special case,
\eqn\scase{\phi={\bar \phi}=\pi/2~,}
the following commutation relations hold:
\eqn\com{[A_0,[A_0,[A_0,A_1]]]={\rm const~}[A_0,A_1]~,\qquad
[A_1,[A_1,[A_1,A_0]]]={\rm const~}[A_1,A_0]~,}
and thus that $A_0$ and $A_1$ are embedded in the larger algebra
\eqn\onsalg{\eqalign{&[A_l,A_m]=4G_{l-m},\cr
&[G_l,A_m]=2A_{m+l}-2A_{m-l},\cr
&[G_l,G_m]=0.}}
This is exactly the same algebra that Onsager~\rons~discovered for the
Ising model in 1944.

The special case~\scase~is now called the superintegrable case. The
discovery that this case satisfies Onsager's algebra was the first time
since Onsager's original paper that a new representation of this algebra
had been discovered. This algebra is the source of many remarkable
simplifications, which make the study of the superintegrable chiral
Potts model particularly interesting~\rbax--\rdkma.

Several years later it was found~\rampty--\rayp~that if the chiral angles
of~\azero~satisfy the restriction that
\eqn\res{\cos\phi=k\cos{\bar\phi}}
then the model~\cp~is integrable it the traditional sense, i.e. it can
be derived from a family of commuting transfer matrices $T(p,q)$:
\eqn\com{[T(p,q),T(p,q')]=0~,}
where $T(p,q)$ is constructed from the Boltzmann weights
\eqn\wv{{W^v_{p,q}(n)\over W^v_{p,q}(0)}=
\prod_{j=1}^{n}\left({d_pb_q-a_pc_q\omega^j\over
b_pd_q-c_pa_q\omega^j}\right)~~{\rm and}~~
{W^h_{p,q}(n)\over W^h_{p,q}(0)}=\prod_{j=1}^n\left({\omega a_p
d_q-d_pa_q\omega^j\over
c_pb_q-b_pc_q\omega^j}\right)}
where $a_p,b_p,c_p,d_p$ and $a_q,b_q,c_q,d_q$ lie on the generalized
elliptic curve
\eqn\curve{a^N+kb^N=k'd^N,\quad ka^N+b^N=k'c^N,\quad k'=(1-k^2)^{1/2}}
which for $k\neq 0,1,\infty$ has genus $N^3-2N^2+1.$
The transfer matrix is defined to be
\eqn\tran{T_{{l},{l'}}(p,q)=\prod_{j=1}^{L}W^v_{p,q}(l_j-l'_j)
W^h_{p,q}(l_j-l'_{j+1})}
with
$l=\{l_1,l_2,\ldots,l_L\}$,
$l_j=0,1,\cdots N-1.$
The Hamiltonian~\cp~is obtained in the limit $p\rightarrow q$ as
\eqn\htran{T(p,q)=1(1+{\rm const}~u)+uH_{\rm CP}+O(u^2)}
where $u$ is a measure of the  deviation of $p$ from $q$ and
the chiral angles $\phi$ and ${\bar \phi}$ satisfy
\eqn\angles{e^{2i\phi/N}=\omega^{1/2}{a_pc_p\over b_p
d_p},\qquad e^{2i{\bar \phi}/N}=\omega^{1/2}{a_pd_p\over b_p c_p}.}

Unfortunately, since the integrability condition~\com~is not as strong as
the condition of Onsager's algebra~\onsalg, the study of this case
is in general
 substantially more difficult than the study of the superintegrable
case~\scase. However there is one exception to this observation;
namely the special case
\eqn\fatzam{\phi={\bar \phi}=0.}
Here, from~\res~, we have $k=1$ and from~\curve~and~\angles~we find
\eqn\rat{a=\omega^{-1/2}b,\qquad c=d~.}
The Boltzmann weights~\wv~reduce to the weights of the model of
Fatteev and Zamolodchikov~\rfz~
\eqn\wfatzam{{W^v(n)\over W^v(0)}=\prod\left(
{\omega^{1/4}e^\lambda-\omega^{j-1/2}\over 1-\omega^{j-1/4}
e^{\lambda}}\right),\qquad {W^h(n)\over
W^h(0)}=\prod_{j=1}^n\left({e^{\lambda}\omega^{j-1/4}-\omega^{-1/2}\over
\omega^j-\omega^{1/4}e^{\lambda}}\right)~,}
where we have set ${a_q/d_q\over a_p/d_p}=\omega^{1/4}e^{\lambda}.$
This model, which for the case $N=3$ is the three-state Potts model, is
solvable by the methods of Bethe's ansatz and conformal field theory.

One of the most interesting properties of the Hamiltonian~\cp~with~\res~is the
spectrum of excitations over the ground state. These excitations can
be studied through the use of the functional equations
satisfied~\ramp\rbs--\rbaxb~by the
transfer matrix of the statistical system~\wv--\tran. In particular
these equations have been used
for the case $N=3$ to study both the superintegrable case~\scase~\rdkma~and
the scalar case~\fatzam~(which is the critical three-state Potts
model)~\radma~\radmb. In the first case, in the massive regime, the
spectrum consists of two quasi-particles. In the second case, in
the ferromagnetic regime, the spectrum has one genuine quasi-particle
and two ``ghost'' particles which carry statistical information but no
energy or momentum.

These two spectra are are not obviously of the same qualitative form.
However, from the phase diagram of the general integrable chiral Potts
model in~\rmr, it is clear that it is possible to
smoothly connect these two points without passing through a massless,
level crossing phase. Thus it should be possible to smoothly connect
the spectra of these two special points together.

It is the purpose of
this paper to study this connection for the case $N=3$. For
convenience, we concentrate on the sector $Q=0,$ where $e^{2 \pi i Q/3}$
are the eigenvalues of the spin rotation operator.
In sec.~2, we summarize the exact results for the ferromagnetic three-state
 Potts and the
superintegrable chiral Potts  model.
In sec.~3, we present the results of a numerical
study of the zeroes of the transfer matrix in the general integrable
case. We conclude in sec.~4 by using this study to connect
the two exact spectra together, and we
contrast our study with previous results~\rvoge--\rhaho.
 Our
principle conclusions are that 1) there is a sense in which the
superintegrable spectrum is more properly regarded as consisting of
three rather than two quasi-particles 2) the momentum ranges for these
three types of quasi-particle are different and 3) as we move smoothly
from the superintegrable to the ferromagnetic three-state Potts point
the allowed momentum range of two of the quasi-particles shrinks to
zero while the range of the third expands to fill the entire Brillouin
zone.

\newsec{Spectrum of the superintegrable chiral
and ferromagnetic three-state Potts models}

The general quasi-particle form for an order one excitation spectrum
over the ground state
is
\eqn\eqpa{\lim_{L\rightarrow \infty}(E_{ex}-E_{GS})=\sum_{\alpha=1}^{N_S}
\sum_{j_{\alpha}=1,{\rm~rules}}^{m_\alpha}e_{\alpha}(P_{j_{\alpha}}^{\alpha})}
and
\eqn\pqpa{\lim_{L\rightarrow
\infty}(P_{ex}-P_{GS})=\sum_{\alpha=1}^{N_S}\sum_{j_{\alpha}=1,{\rm~
rules}}^{m_{\alpha}}P_{j_{\alpha}}^{\alpha}}
where $N_S$ is the number of species of quasi-particles, $m_{\alpha}$
is the number of quasi-particles of type $\alpha$ in the states,
$e_{\alpha}(P)$ are the single particle energies, and the subscript
``rules'' indicates the restriction on the allowed choices for
$P_{j_{\alpha}}^{\alpha}.$ If one of the rules is
\eqn\fermi{P_{j_{\alpha}}^{\alpha}\neq P_{k_{\alpha}}^{\alpha}~~{\rm
for}~~j_{\alpha}\neq k_{\alpha}}
then the quasi-particles are said to be fermionic.

The spectrum of the ferromagnetic three-state Potts chain has been
computed~{\radma~\radmb} in terms of the zeroes
$\lambda_j$ of the eigenvalues $\Lambda$ of the transfer matrix.
\eqn\eigzer{\Lambda=\left({ \sinh(\pi i/6)\sinh(\pi
i/3)\over\sinh{1\over2}({\pi i\over2}-\lambda)\sinh{1\over2}
({\pi i\over2}+\lambda)}\right)^L\prod_{k=1}^{2L}
{\sinh{1\over2}(\lambda-\lambda_k)\over\sinh{1\over2}
({\pi i\over6}+\lambda_k)}~,}
as
\eqn\bethener{E=\sum_{k=1}^{2L}\cot{1\over 2}(i
\lambda_k+{\pi\over6})-
{2L\over3^{1/2}}~,}
where $\lambda _j$ satisfy the Bethe equation
\eqn\beteqn{(-1)^{L+1}\left( {\sinh{1\over 2}(\lambda_j-2iS\gamma)
\over\sinh{1\over2}(\lambda_j+2iS\gamma)}\right)^{2L}=
\prod_{k=1}^{M}{\sinh {1\over2}(\lambda_j-\lambda_k-2i\gamma)
\over\sinh{1\over2}(\lambda_j-\lambda_k+2i\gamma)}~,}
with $S=1/4$, $\gamma=\pi/3$ and (for $Q=0$)  $M=2L$
{}From this it
is found that as $L\rightarrow \infty$  the allowed imaginary parts
 of  $\lambda_j$
are
\eqn\imlamb{{\rm Im~}\lambda_j=0,~\pi,~\pm\pi/3,~\pm2\pi/3,~\pm\pi/2}
which we refer to as $+, -, 2s, -2s$ and $ns$ respectively.
We find that
the spectrum is of the fermionic quasi-particle
form, where there is only one type of single particle energy. In particular
\eqn\pottsen{\lim_{L\rightarrow\infty}
(E(\{P_j\})-E_{GS})=\sum_{j=1}^{m_+}e_+(P^+_j)}
where
\eqn\pottsin{e_{+}(P)=6\sin(P/2),~~~0\leq P\leq 2\pi,}
and each energy occurs with a multiplicity
\eqn\mult{{{1\over 3}(m_+-2m_{ns}-m_{-2s})\choose
m_{-2s}}{{1\over 3}(2m_+-m_{ns}-2m_{-2s})\choose m_{ns}}~,}
where the integers $m_+,m_{ns}$ and $m_{-2s}$ take on all nonnegative
values subject only to the restriction (valid for the sector $Q=0$
under consideration)
\eqn\mrestr{m_+-2m_{ns}-m_{-2s}\equiv 0~({\rm mod}~3)~.}
This degeneracy factor can be interpreted as indicating that the $ns$
and $-2s$ excitations  carry zero
momentum and energy. We refer to these zero energy excitations as
``ghost'' excitations.

More precisely~\rdkmm, the spectrum
to order $1/L$
is given by
\eqn\etoonel{E(\{P\})-E_{GS}=
\sum_{\alpha=+,ns,-2s}\sum_{j_\alpha}^{m_\alpha}e_{\alpha}
(P_{j_\alpha}^{\alpha})~,}
where the energies of the ``ghost'' excitations are
\eqn\enghost{e_{ns,-2s}(P_{j_\alpha}^{\alpha})=3P_{j_\alpha}^{\alpha}~,}
and the momenta $P_{j_{\alpha}}^{\alpha}$ obey the Fermi
exclusion
rule~\fermi, and
 are chosen from sets with
spacings $2\pi/L$ with the following limits
\eqn\pselection{\eqalign{-{\pi\over L}\left[
{1\over3}(m_+-2m_{ns}-m_{-2s})-1\right]& \leq P^+_j
\leq 2\pi+{\pi\over L}\left[ {1\over 3}(m_+-2m_{ns}-m_{-2s})-1\right]~,\cr
-{\pi\over L}\left[({1\over 3}(m_+-2m_{ns}-m_{-2s})-1\right]& \leq
P^{-2s}_j\leq{\pi\over L}\left[{1\over 3}(m_+-2m_{ns}-m_{-2s})-1\right]~,\cr
-{\pi\over L}\left[{1\over 3}(2 m_+-m_{ns}-2m_{-2s})-1\right]& \leq
P^{ns}_j \leq {\pi\over L}\left[{1\over 3}(2m_+-m_{ns}-2m_{-2s})-1\right]~.}}
We refer to such momentum exclusion rules as generalized Fermi statistics.
It is important to note that in ~\pselection~  only the $+$
momentum take on a macroscopic number of values (proportional to
$L$). The $-2s$ and the $ns$ ``ghost'' momenta take on only a finite
number of values (even as $L\rightarrow \infty$). We refer to such a
momentum range as microscopic.

The energy spectrum of the superintegrable chiral Potts model has also
been computed. At this point we find from~\angles~that
\eqn\siabcd{a_p=b_p,\qquad c_p=d_p}
and hence
\eqn\aoverd{a_p/d_p=\eta^{-1},~~{\rm where}~~\eta=[(1+k)/(1-k)]^{1/6}.}
{}From the property of superintegrability it
follows~\rdav~that the
spectrum of eigenvalues is decomposed into a number of sets of the form~\rampt
\eqn\supespec{E=A+kB+3\sum_{j=1}^{m_E}{\pm}(1+k^2+a_jk)^{1/2}}
where $A,~B,~a_j$ and the number of terms $m_E$ varies from set to
set. More precisely we find~{\ramp~\rdkma} that
the eigenvalues of the transfer matrix are of the form
\eqn\sitraeig{\eqalign{\Lambda_{si}&={3^L(\eta a/d-1)^L\over[(\eta a/d)^3-1]^L}
(\eta {a\over d})^{P_a}(\eta {b\over c})^{P_b}({c^3\over d^3})^{P_c}\cr
&\prod_{l=1}^{m_P}\left( {1+\omega v_l\eta^2 ab/cd\over 1+\omega v_l}\right)
\prod_{l=1}^{m_E}\left({1+k\over1-k}\right)^{1/2}\left({a^3+b^3\over
2d^3}
\pm w_l{a^3-b^3\over (1+k)d^3}\right)~,}}
where we have used the sign convention for $v_l$ of~\ramp.
These $v_l$ satisfy
\eqn\sibethe{\left(\omega^2+v_k\over
\omega+ v_k \right)^L=-\omega^{-(P_a+P_b)}
\prod_{l=1}^{m_P}\left({v_k-\omega^2v_l\over v_k-\omega v_l}\right)~,}
The $w_l$ in~\sitraeig~are
\eqn\well{w_l^2={1\over 4}(1-k)^2+{k \over 1-t_l^3}}
where the $t_l$ are the roots of the polynomial
\eqn\poly{\eqalign{P(t)&=t^{-(P_a+P_b)}\Bigl[ (\omega^2 t-1)^L(\omega
t-1)^L\omega^{P_a+P_b}\prod_{l=1}^{m_P}\left( {1+tv_l\over
1+t^3v^3_l}\right)\cr
+&(t-1)^L(\omega^2t-1)^L\prod_{l=1}^{m_P}
\left({1+\omega tv_l\over 1+t^3v_l^3}\right)
+(t-1)^L(\omega t
-1)^L\omega^{-(P_a+P_b)}\prod_{l=1}^{m_P}\left( {1+\omega^2 t
v_l\over 1+t^3 v_l^3}\right) \Bigr] }}
and $P_a$, $P_b$ and $P_c$ are integers.
The eigenvalues of the superintegrable chain are thus
\eqn\eigsi{E_{\rm SI}=A+Bk+6\sum_{l=1}^{m_E}\pm w_l.}
In the small $k$ regime where there is a mass gap the ground state
corresponds to the choice $m_P=0$ in~\poly~and all  minus signs in~\eigsi.

To obtain the excitation spectrum we need to know the allowed
solutions $\lambda_l$ of the Bethe's equation~\sibethe. This was
studied in~\rdkma~where it was found that as
$L\rightarrow \infty$ there are three
allowed values for the imaginary parts of $\lambda_j$
\eqn\silam{{\rm Im}\lambda_j=0,~\pi,~\pm 2\pi/3,}
which we call $+,~-,$ and $-2s$ respectively. These correspond to
\eqn\vval{v_l=v^+>0,~v^-<0~{\rm and}~v^{-2s}e^{\pm 2\pi/3}.}
It was also shown that the eigenvalues obtained from the choices of
the $\pm$ signs in~\eigsi~can be written in terms of the excitations
$+$ and $-2s$
of the Bethe's equation with a suitable alteration of the rules of
combination. Thus it was shown that
if we use the relation between momenta and $v_l^{\alpha}$ of
\eqn\momv{e^{-iP^{r}}=\left({1+\omega^2 v^{r}\over 1+\omega
v^{r}}\right)~~{\rm for}~~r=\pm~,\qquad
e^{-iP^{-2s}}=\left( {1+e^{-\pi i/3}v^{-2s}\over 1+e^{\pi
i/3}v^{-2s}} \right)~, }
then the order one
excitations are of the fermionic quasi-particle form~\eqpa~where the
single particle energies are
\eqn\sien{ {e_r}(P^r)=2|1-k|+{3\over \pi}\int_{1}^{|{1+k\over 1-k}|^{2/3}}
dt\left({\omega v^r\over \omega t v^r+1}+{\omega^2 v^r\over \omega^2 t
v^r+1}\right)\left[ {4 k\over t^3-1}-(1-k)^2\right]^{1/2}~,}
where $r$ may be either $\pm$, with the range of $P^r$ restricted to
\eqn\realp{0\leq P^+ \leq {4\pi\over 3},\qquad{4\pi\over 3}
\leq P^-\leq 2\pi~,}
and
\eqn\sientwos{e_{-2s}(P^{-2s})=4|1-k|+
{3\over \pi}\int_1^{|{1+k\over 1-k}|^{2/3}}
dt{v^{-2s}[4(v^{-2s}t)^2+v^{-2s}t+1]\over (v^{-2s}t)^3-1}\left[{4k\over
t^3-1}-(1-k)^2\right]^{1/2}~,}
with
\eqn\twosp{{2\pi\over 3 }\leq P^{-2s}\leq 2\pi~.}
We also have the restriction
\eqn\mressi{m_{+}+m_{-}+2m_{-2s}\equiv 0~({\rm mod}~3).}

Moreover this result may be extended
to  order $1/L$ by use of the counting rules of~\rdkma, where we
note that the rewriting of the excitations associated with the $\pm$
signs in~\eigsi~in terms of $+$ and $-2s$ excitations is possible,
 because any set of
roots $v_l=v,~\omega v~{\rm and}~\omega^2 v$ satisfy
{}~\sibethe~independently of the value of $v.$
Thus the reduction in the number of states which results from choosing
all the $\pm$ signs to be negative is exactly compensated for by
elimination of the exclusion rule $-v^{-}_k\neq v^{-2s}_l$ of ref.~\rdkma.
Using this we find, to order $1/L$ (for $L\equiv0~({\rm mod~}3)$ where
$P^a=P^b=0$),
\eqn\esioonel{E(\{P\})-E_{GS}=\sum_{\alpha=+,-,-2s}\sum_{j_\alpha}^{m_\alpha}
e_{\alpha}(P^{\alpha}_{j_{\alpha}})~,}
where the momenta $P^{\alpha}_{j_\alpha}$ obey the Fermi
rule~\fermi~and
are chosen from sets with spacings $2\pi/L$ with the following
limits
\eqn\simomlim{\eqalign{-{\pi \over L}\left[{1\over
3}(m_{+}-2m_{-}-m_{-2s})-1\right] &\leq P_j^{+} \leq{4\pi\over
3}+{\pi\over L}\left[{1\over3}(m_{+}-2m_{-}-m_{-2s})-1\right],\cr
{4\pi\over 3}-{\pi \over L}\left[{1\over
3}(2m_{+}-m_{-}-2m_{-2s})-1\right] &\leq P_j^- \leq 2\pi+{\pi\over
L}\left[{1\over 3}(2 m_{+}-m_{-}-2m_{-2s})-1\right],\cr
{2\pi\over 3}-{\pi\over L}\left[{1\over
3}(m_{+}-2m_{-}-m_{-2s})-1\right] &\leq P_j^{-2s} \leq 2\pi+{\pi\over
L}\left[{1\over 3}(m_{+}-2m_{-}-m_{-2s})-1\right].}}
Note that
the $m_{\alpha}$ dependence of these restrictions is identical with
that of the restrictions~\pselection~of the ferromagnetic three-state
Potts model if we make the identification:
\eqn\ident{(+,-,-2s)~{\rm superintegrable}~=(+,ns,-2s)~{\rm
ferromagnetic~three-state~Potts}.}
Moreover this identification makes the restrictions{~{\mrestr}}
and~\mressi~identical. The only differences between
the two cases are 1) the single particle
energy functions are different, 2) the macroscopic momentum ranges
of the $+$ excitation is $2\pi$ in~\pselection~ and is $4\pi/3$ in
{}~\simomlim~and 3) the momentum ranges of the $-2s$ and
the $ns$ excitations are microscopic in ~\pselection~ whereas the ranges of
$-$ and $-2s$ excitations in~\simomlim~ are macroscopic. We conjecture that
in the general case that the spectrum is of the form of the
superintegrable spectrum where the single particle energies and the
macroscopic momentum ranges depend on the chiral angles.

\newsec{Spectrum in the general case}

We now turn to an investigation of the excitation spectrum of the general
integrable chiral Potts model and see what evidence can be obtained to
support the conjecture of the previous section. An initial
investigation of this spectrum was made in~\rmr~by use of the
functional equations of~\rbbp~and~\rbaxb. The single particle energy
function was found to be
\eqn\genep{\eqalign{e(v)&=2(1-k)t_p^{3/2}\pm3^{1/2}vt_p^{-1/2}
{\{(t_p^3-1)[(1+k)^2-(1-k)^2t_p^3]\}^{1/2}
\over(\omega t_pv+1)(\omega^2t_pv+1)}\cr
&+{3vt_p^{3/2}\over \pi}P\int_1^{|{1+k\over 1-k}|^{2/3}}dt\left({\omega\over
\omega tv+1}+{\omega^2\over
\omega^2 tv+1}\right){\{(t^3-1)[(1+k)^2-(1-k)^2t^3]\}^{1/2}\over t^3-t_p^3},}}
where the $\pm$ sign is chosen to be $+$ if $0\leq\phi\leq\pi/2$ and $-$ if
$\pi/2 \leq \phi\leq \pi$ , $P$ indicates the principle value of the
integral, $t_p$ satisfies $1\leq t_p\leq ({1+k\over 1-k})^{2/3}$ and is
defined to be
\eqn\tp{t_p^3={(1+k)^2\over 1+k^2(1-2\cos^2{\bar \phi})+2k|\sin{\bar
\phi}|(1-k^2\cos^2{\bar \phi})^{1/2}}}
and the allowed values of $v^{-1}$ are equal to
$\omega^{1/2}$ times the location of
 the zeroes of the eigenvalues of the
transfer matrix which satisfy an equation ((42) of~\rmr~) which resembles a
Bethe's equation. However, this equation for the allowed $v$ is in
fact not a Bethe's equation like~\beteqn~and we have not been able
to relate the solutions $v_j$ to the momentum of the state.
Consequently we do not have an analytic verification of the
conjecture.

To proceed further we have made a numerical study of the zeroes of the
transfer matrix. For convenience we consider the case $k=1$. Here we
define ${\bar d}^3=[(1-k)/(1+k)]^{1/2}d^3$ and find that the
curve~\curve~reduces to the Fermat curve
\eqn\fermat{a^3+b^3=2{\bar d}^3~,}
and that the superintegrable eigenvalue~\sitraeig~reduces to
\eqn\fereig{\Lambda={3^L(a/{\bar d}-1)^L\over [(a/{\bar
d})^3-1]^L}({a\over {\bar d}})^{P_a} ({a\over {\bar d}})^{P_b}
\prod_{l=1}^{m_P}\left( {1+\omega v_l ab/{\bar d}^2\over 1+\omega
v_l}\right)\prod_{l=1}^{m_E}
{( 1\pm w_l({a^3-b^3\over 2{\bar d}^3}))}~.}
For our purposes we rewrite this in terms of the variables
\eqn\tbar{{\bar t}=\omega^{-1/2}{ab\over {\bar d}^2},\qquad
u={a^3-b^3\over 2 {\bar d}^3}}
as
\eqn\fereigtwo{\Lambda=
{3^L({a/ {\bar d}}-1)^L\over [(a/ {\bar
d})^3-1]^L}\left({a\over {\bar d}}\right)^{P_a}\left({b\over {\bar
d}}\right)^{P_b} \prod_{l=1}^{m_p}\left({1- v_l {\bar t}\over
1+\omega v_l}\right) \prod_{l=1}^{m_E}(1- w_l u)~,}
where from~\fermat~${\bar t}$ and $u$  satisfy
\eqn\tueqn{u^2=1+{\bar t}^3.}
For arbitrary values of the chiral angle $\phi$,
the eigenvalues of the transfer matrix
are still meromorphic functions on the Riemann surface~\tueqn~even
though the form~\fereigtwo~is no longer valid.

In the superintegrable case we see from~\fereigtwo~that there are
qualitatively two different types of zeros. The first type is
specified by a value of ${\bar t}_l$, which satisfies
\eqn\tveqn{{\bar t}_l=v_l^{-1}}
and exists for both sheets $(\pm)$ of $u$. The second type exists only on
one sheet of $u$ (either $+$ or $-$) and has three values of ${\bar t}_l$
related by
\eqn\uroots{{\bar t}^{(2)}_l=\omega {\bar t}^{(1)}_l,~~~
{\bar t}^{(3)}=\omega^2{\bar t}_l^{(1)}}
where ${\bar t}_l^{(1)}$ is real.

We have studied the zeroes of the transfer matrix for arbitrary values
of the chiral angle $\phi={\bar \phi}$ by use of the procedure
previously used for three-state Potts
case~\radma~. We will here illustrate the behavior of the zeroes by
presenting the results for the nine eigenvalues in the $Q=0$
sector for $L=3$.
This behavior is illustrated
schematically in
figs. 1--9. We summarize some of the features of these results in
table 1.

There are many features of the motion of the zeroes which may be seen
in these figures and we will explicitly comment on only a few of them.
First, all zeroes move towards infinity as $\phi \rightarrow 0.$
Second, we note that no zero ever moves from one sheet of $u$ to the other, so
the sign specifying the sheet in the superintegrable case acts like a
good quantum number. Third, we see that as $\phi\rightarrow 0$ most
of the zeroes stay close to the rays in the ${\bar t}$ plane ${\rm
arg}~{\bar t}=0,\pm 2\pi/3.$ These correspond to the roots $\pm$ and $\pm 2s$
of the three-state Potts model. However there are some roots, as
seen in Figs.~2 and 6, which collide with the branch point at ${\bar
t}^3=-1$ and move to infinity on somewhat less well defined rays. These
are the $ns$ roots of the three-state Potts model whose imaginary
parts are subject to much more deviation from the asymptotic value
than the other roots.

$$\vbox{\tabskip=0pt \offinterlineskip
\def\tablerule{\noalign{\hrule}}
\halign to305.81876pt{\vrule#&\strut$~#~$&#\vrule&\strut$~#~$&#
\vrule&\strut$~#~$&#\vrule&\strut$~#~$&#\vrule&\strut$~#~$&#\vrule&
\strut$~#~$&#\vrule&\strut$~#~$&#\vrule&\strut$~#~$&#
\vrule&\strut$~#~$&#\vrule\cr\tablerule
&{\rm fig.}&&P&&m_E&&+&&-&&-2s&&E_{\rm SI}&&{\rm 3sP}~{\rm
content}&&E_{\rm 3sP}&\cr\tablerule
&1&&0&&-~-&&0&&0&&0&&-7.7459&&~~~~3(2s)&&-8.87348&\cr
&2&&0&&+~-&&0&&0&&0&&-3.4641&&(2s)~(ns)~2(+)&&-3.4641&\cr
&3&&0&&{\rm none}&&3&&0&&0&&0.0&&(2s)~(-)~3(+)&&~3.4641&\cr
&4&&0&&-~+&&0&&0&&0&&~3.4641&&(2s)~(-)~3(+)&&~3.4641&\cr
&5&&0&&+~+&&0&&0&&0&&~7.7454&&~~~(-2s)~4(+)&&5.40938&\cr
&6&&-2\pi/3&&{\rm none}&&1&&0&&1&&0.0&&(2s)~(ns)~2(+)&&0.0&\cr
&7&&-2\pi/3&&{\rm none}&&2&&1&&0&&0.0&&(2s)~(-)~3(+)&&0.0&\cr
&8&&~2\pi/3&&{\rm none}&&1&&0&&1&&0.0&&(2s)~(-)~3(+)&&0.0&\cr
&9&&~2\pi/3&&{\rm none}&&2&&1&&0&&0.0&&(2s)~(ns)~2(+)&&0.0&\cr\tablerule}}$$
{{\bf Table~1.} A description of the zeroes of the transfer
matrix studies in figures 1--9. The zeroes are indicated by their
position in the ${\bar t}$ plane of~\tbar~and the sign of $u.$
We indicate in the columns $m_E$ $+,-,-2s$ and $E_{\rm SI}$ the content
and energy of the superintegrable case where $\phi=\pi/2$.
The
first (second) sign in the column $m_E$ refers to the root with the
largest (smallest) value of ${\bar t}.$ We indicate in the columns 3sP
content and $E_{\rm 3sP}$ the content and energy of the three-state Potts case
where $\phi=0$.}

\bigskip

The most obvious property of the motion of these zeroes is that the
zeroes change their character.

In Fig.~1 the positive real zeroes for $u$ negative move to the rays
${\rm arg~}{\bar t}=\pm2\pi/3.$ This corresponds to the motion of the
zeroes in the ground state of the massive phase.

In Fig.~2 the $u<0$ roots on the rays ${\rm arg}~{\bar t}=\pm 2\pi/3$ stay
near the ray but the $u>0$ roots that stay on ${\rm arg}~{\bar
t}=\pm 2\pi/3$ execute a motion that has one of them pairing with a $u<0$
root to become an $ns$ pair as $\phi\rightarrow 0$. The remaining $u>0$
root moves out along the positive real axis to become a + root of the
three-state Potts model.

Indeed in all the remaining figures there are several roots which
undergo a qualitative change in going from the superintegrable case to
the three-state Potts case. These qualitative changes must be
accounted for in the solution of the pseudo-Bethe equation of~\rmr.

\bigskip

\newsec{Conclusions}

There are several points to consider before the data
presented in figures 1--9
can be used to discuss the
conjecture presented at the end of sec.~2.

First, since the ground state of the superintegrable chiral
Potts model at $k=1$ is not the state with $m_+=m_-=m_{-2s}=0,$ the
identification of the ground state configuration of zeroes made
in the previous section is not literally correct for the
$L\rightarrow \infty$ limit. Indeed, the chiral Potts model is
massless all along the line $k=1.$ However, for systems of such small
size as $L=3$ the level crossing phenomena which causes
$m_+=m_-=m_{-2s}$ not to be the ground state at $k=1$ cannot be seen
(in fact it is found that L must be larger than
18 for level crossing to occur in the
superintegrable case). Moreover we have made further numerical studies
of the motion of the zeroes starting from a value of $k$ sufficiently
small that the superintegrable case is massive and followed a path in
$({\bar \phi},k)$ space that always lies in the massive phase.
The only qualitative difference is that there is an additional
square root branch point at ${\bar t}^3=-[(1+k)/(1-k)]^2$ and this
branch point does not cause a qualitative change in the motions of the
zeroes in the figures 1-9.

Secondly, we must consider how much physical intuition should be
invested in the zeroes of the transfer matrix. For the three-state
Potts model  and for the $r=6$ RSOS model (which is related to the
three-state Potts model by an orbifold construction~\rfg~\rpz)
there is an alternative set of Bethe's
equations~\rbaza--\rkun~which
focus on the zeroes of an auxiliary objet called $Q.$
This formalism is widely used to compute spectra of integrable models.
For the
general chiral Potts models there are many different
functional equations~\rbs\rbbp~and no
one seems to be preferred for the purpose of giving physical
interpretations.
We use the $T$-matrix approach here for our convenience.

A major question of physical interest is the
the uniqueness (or lack thereof) of the quasi-particle description
of the spectrum.
This problem is seen vividly if we compare our conjecture of sec.~2 with
the numerical studies and interpretation of~\rvoge\rhaho. In these
studies the spectrum of the massless ferromagnetic three-state Potts
model is interpreted as a quasi-particle spectrum made up of two
excitations .
This interpretation is completely consistent with the spectrum of the
(massless) hard hexagon
model (r=5 RSOS at the I/II boundary)~\rhexa--\rhexc~which is in
the same universality class as the ferromagnetic
three-state Potts model. Moreover,
the hard hexagon model remains integrable in the massive (regime II) phase
(although the three-state Potts model does not) and the two
quasi-particle interpretation describes this massive
spectrum~\rhexa. In ref.~\rvoge, where the case $\phi={\bar
\phi}$ (which is integrable only for
$k=1,~\phi={\bar \phi}=0$ and ${\phi={\bar \phi}=\pi/2}$) is
extensively studied, this two quasi-particle picture is used to
interpret the numerical results.

It would appear that the two quasi-particle description of the
spectrum in the case $\phi={\bar \phi}=0,~k=1$ is not the same as
the description in terms of one quasi-particle and two ghosts found
in~\radma\radmb. In order to justify our conjecture we must address
this apparent contradiction. This question has been extensively studied at
the level of the order $1/L$ excitations  in~\rdkmm.
The two quasi-particle form
of the spectrum leads to the $q$-series character formula of Lepowsky
and Primc~\rlepr~and the spectrum of sec.~2 leads to a different form
which involves $q$-binomial coefficients~\rdkmm. However there is an
identity on $q$-series~\rberk~which makes these two different
looking fermionic forms of the character equal. If the study
of~\rvoge-\rhaho~is a correct interpretation of the numerical data
then this identity of characters must have an extension to the order
one spectrum.
Let us
assume such an identity and examine
its
consequences.

The existence of alternative forms of a massless order one spectrum
was first seen in the Ising model where the spectrum of the transfer
matrix at
the critical temperature
$T_c$ can be obtained either in terms of an odd number of
quasi-particles if  $T_c$ is approached from above, or an even
number of quasi particles if $T_c$ is approached from below.
Spectra which can involve ghosts
have been investigated in detail in the spin 1/2 Heisenberg
antiferromagnet by Faddeev and Takhtajan~\rfadtak~\rfadtakb~, who showed the
equivalence of a two quasi-particle spectrum with the previously known
spectrum~\rtaksuz~\rgau~ of one
quasi-particle and an infinite number of ghosts. In the first case the
two quasi-particles are interpreted~\rfadtak~\rfadtakb~
as spin $1/2$ spin waves.
In the second case~\rtaksuz~\rgau~ the quasi-particles are the holes in the
ground state
distribution of roots of the Bethe's equation and the ghosts
(string solutions of the Bethe equations)
 which have
zero energy and zero momentum are  similar to the ghosts seen in the
massive Thirring model~\rkor. The quasi-particles
in~\rfadtak-\rgau~are the interacting counterparts of the free spinons
of the $1/r^2$ Heisenberg spin chain~\rhal-\rha.The corresponding
$1/L$ spectrum of the related conformal field theory has been
investigated in ~\rhalb~and ~\rbps. The character identity between
the two representations of the spectrum was proven in ~\rsch.

However it must be
explicitly stated that the existence of an identity between different
quasi-particle  representations of the spectrum forces us to
question what we mean by the ``physical reality''
of a quasi-particle description (at least for a massless spectrum).
We do not seem to have the right to
ascribe physical reality to something which is not unique. Indeed it
is only the spectrum of the Hamiltonian which has a unique meaning.
Any quasi-particle description of this spectrum imposes a basis on the
Hilbert space and it is this basis to which we give the physical words
of quasi-particle. But logically speaking a basis dependent statement
can only be given physical reality if an additional  condition has been
given, which singles out the particular basis to be used. In the absence of
such a condition different bases cannot be distinguished and the non
uniqueness of the description must be accepted. Moreover we point out
the fact that many characters have been found to have more that one
fermionic representation~\rchara-\rcharb.
This seems to imply
that at massless
points of high symmetry there may be several different quasi-particle
interpretations of exactly the same spectrum. This is the infinite
dimensional analogy of finite dimensional degenerate perturbation theory,
where different bases are used depending on the type of perturbation applied.
The basis used here is appropriate to the chiral interaction which
breaks the $Z_2$ charge conjugation symmetry of $S_3$ symmetric
ferromagnetic Potts model. This basis also seems appropriate~\rdkmm~for
 the integrable perturbation of the critical three-state Potts
model which breaks $Z_3$ symmetry~\rkm~\rpas.
The basis with two quasi-particles with
equal energies for $Q=\pm 1$ seems to be appropriate for a (non-integrable)
perturbation which
preserves the full $S_3$ symmetry.

\bigskip

With this discussion of the symmetry of the ferromagnetic massless
three-state Potts point we may now proceed to use the data of sec.~3
to discuss the conjecture of sec.~2. There are indeed two excitation
energies: $e_{\pm}(P)$ has $Q=+1$ and $e_{-2s}(P)$ has $Q=-1.$ The
$-2s$ excitations can be thought of, in some sense,
 as bound states of the $\pm$
excitations and the motion of the zeroes with $u\geq0$
from the rays $\arg {\bar
t}=\pm2\pi/3$ to the real axis can be thought of as  an unbinding
transition. As $\phi \rightarrow 0$ the number of these $-2s$ excitations
decreases until at $\phi={\bar \phi}=0$ the number becomes
microscopic (as is needed to reproduce the ~\pselection~).
Moreover there is some difference between the $+$ and the $-$
excitations. Taken together they span the entire Brillouin zone $0\leq
P\leq 2\pi$ and as a function of the order one momentum the energy is
continuous in this Brillouin zone. But nevertheless there is a break in
the counting rules for the states, which separates the $+$ from the $-$
excitations.
Some of the $-$ excitations move smoothly to the $ns$ excitations, which
also must become microscopic in the limit $\phi={\bar \phi}=0$ in
order to agree with ~\pselection. Of course with such small systems it
is hardly possible to see these shifts from macroscopic to microscopic
in any quantitative way. Nevertheless the existing data supports this
picture in a qualitative fashion and it is
in this sense that we say that the data of section 3 supports the
conjecture of sec.~2.

It is sometimes said that the massless conformal field theories cannot
be given a particle interpretation. This statement is based, in part,
on the notion of quasi-particle as an isolated pole in a Greens
function and, obviously, if massless particles are present there are
no isolated poles. However, for genuine physical applications this
notion is not particularly useful, because  massless excitations such
as phonons and magnons are common in condensed matter physics and,
moreover, all charged particles couple
to massless photons. Indeed, we are used to the fact that different
gauges may differ in their description of the ``longitudinal
photons.''
In QED this non uniqueness is not considered to be a problem and it is
often said that  the Coulomb gauge is singled out as the one which
should be considered as being physical.
However, in the distinction between the various massive perturbations
of the ferromagnetic Potts model we seem to be seeing cases where
several different bases have physical relevance. It thus seems to us
that it is more appropriate to say that conformal field theories not
only do have a quasi-particle description but that they may in general have
several different descriptions. The notion of perturbation of
conformal field theory thus is related to the question of how many of
such quasi-particle descriptions can be found.

{\bf Acknowledgements}
We are pleased to acknowledge useful discussions with
Prof. A.W.W. Ludwig and Prof. K. Schoutens.
This work is partially supported by the National Science Foundation
under grant DMR-9106648.

{\bf Figure captions}

{\bf Figure 1} Schematic plot of the motion of the zeroes of the eigenvalue
with $P=0,~Q=0$ whose superintegrable
content is $m_E=-~-,~m_+=m_-=m_{-2s}
=0.$ The arrows indicate the motion of the zeroes as $\phi$ starts
from $\pi/2$ and moves towards zero. The two zeroes which are
initially on the real axis remain on the real axis until they collide
at ${\bar t}=0$ when $\phi=\pi/3.$ We schematically represent this
situation by lines which are off the real axis as a means of
visualization. This convention for the motion of zeroes on the real
axis will be used in all figures.

{\bf Figure 2} Schematic plot of the motion of the zeroes of the eigenvalue
with $P=0,~Q=0$ whose superintegrable content is
$m_E=+~-,~m_+=m_-=m_{-2s}=0.$

{\bf Figure 3} Schematic plot of the motion of the zeroes of the eigenvalue
with
$P=0,~Q=0$ whose superintegrable content is $m_+=3,m_-=m_{-2s}=0.$

{\bf Figure 4} Schematic plot of the motion of the zeroes of the eigenvalue
with $P=0,~Q=0$ whose superintegrable content is
$m_E=-~+,~m_+=m_-=m_{-2s}=0.$

{\bf Figure 5} Schematic plot of the motion of the zeroes of the eigenvalue
with $P=0,~Q=0$ whose superintegrable content is
$m_E=+~+,~m_-=m_+=m_{-2s}=0.$

{\bf Figure 6} Schematic plot of the motion of the zeroes of the eigenvalue
with $P=-2\pi/3,~Q=0$ whose superintegrable content is $m_+=m_{-2s}=1$
and $m_-=0.$

{\bf Figure 7} Schematic plot of the motion of the zeroes of the eigenvalue
with $P=-2\pi/3,~Q=0$ whose superintegrable content is
$m_+=2,~m_-=1,~m_{-2s}=0.$

{\bf Figure 8} Schematic plot of the motion of the zeroes of the eigenvalue
with $P=2\pi/3,~Q=0$ whose superintegrable content is $m_+=m_{-2s}=1$
and $m_-=0.$

{\bf Figure 9} Schematic plot of the motion of the zeroes of the eigenvalue
with $P=2\pi/3,~Q=0$ whose superintegrable content is
$m_+=2,~m_-=1,~m_{-2s}=0.$
\smallskip

\vfill

\eject
\listrefs

\vfill\eject

\bye
\end